\documentclass[%
 reprint,
superscriptaddress,
 twocolumn,
 sort&compress,
 amsmath,amssymb,
 aps,
 prx,
]{revtex4-2}

\usepackage[usenames,dvipsnames]{xcolor}
\usepackage{amssymb}
\usepackage{amsmath}
\usepackage{dsfont}
\usepackage{txfonts}

\usepackage{graphicx}
\usepackage{mathptmx, textcomp, float}
\usepackage{braket,amsfonts}
\usepackage[dvipsnames]{xcolor}
\usepackage{upgreek}
\usepackage{txfonts}
\usepackage{braket}
\usepackage{babel}
\usepackage{blindtext}
\usepackage{physics}
\usepackage{enumitem}
\usepackage[normalem]{ulem}
\usepackage[hidelinks,hypertexnames=false]{hyperref}
\usepackage[all]{hypcap}

\usepackage{rotating}

\setlist[itemize]{itemsep=0pt}

\renewcommand{\Im}{\text{Im}}
\renewcommand{\Re}{\text{Re}}

\newcommand{\JJ}{{J}}
\newcommand{\brket}[1]{\langle {#1} \rangle }

\begin{document}

\title{Stability of superfluids in tilted optical lattices with periodic driving}

\author{Robbie Cruickshank}
\affiliation{Department of Physics and SUPA, University of Strathclyde, Glasgow G4 0NG, United Kingdom }
\author{Andrea Di Carli}
\affiliation{Department of Physics and SUPA, University of Strathclyde, Glasgow G4 0NG, United Kingdom }
\author{Matthew Mitchell}
\affiliation{Department of Physics and SUPA, University of Strathclyde, Glasgow G4 0NG, United Kingdom }
\author{Arthur~La~Rooij}
\affiliation{Department of Physics and SUPA, University of Strathclyde, Glasgow G4 0NG, United Kingdom }
\author{Stefan Kuhr}
\affiliation{Department of Physics and SUPA, University of Strathclyde, Glasgow G4 0NG, United Kingdom }
\author{Charles E. Creffield}
\affiliation{Departamento de F\'{i}sica de Materiales, Universidad Complutense de Madrid,
E-28040 Madrid, Spain}
\author{Elmar Haller}
\affiliation{Department of Physics and SUPA, University of Strathclyde, Glasgow G4 0NG, United Kingdom }

\date{\today}

\begin{abstract}
Tilted lattice potentials with periodic driving play a crucial role in the study of artificial gauge fields and topological phases with ultracold quantum gases. However, driving-induced heating and the growth of phonon modes restrict their use for probing interacting many-body states. Here, we experimentally investigate phonon modes and interaction-driven instabilities of superfluids in the lowest band of a shaken optical lattice. We identify stable and unstable parameter regions and provide a general resonance condition. In contrast to the high-frequency approximation of a Floquet description, we use the superfluids' micromotion to analyze the growth of phonon modes from slow to fast driving frequencies. Our observations enable the prediction of stable parameter regimes for quantum-simulation experiments aimed at studying driven systems with strong interactions over extended time scales.
\end{abstract}
\maketitle

\section{Introduction}

Periodic driving of quantum gases in optical lattices provides versatile mechanisms to create tailored lattice potentials for quantum simulation experiments \cite{goldman2014b, eckardt2017}. Recently, periodic driving facilitated the realization of artificial magnetic fields \cite{aidelsburger2011,struck2012,parker2013} and topological lattice models \cite{aidelsburger2018b,cooper2019}, such as the Harper–Hofstadter and Haldane models \cite{aidelsburger2013b,miyake2013c,jotzu2014a}. Furthermore, periodic driving has been used to study quantum phase transitions \cite{zenesini2009,clark2016d} and quantum critical points \cite{feng2018}, to create ``fireworks'' and patterns in Bose-Einstein condensates \cite{clark2017a,zhang2020a} and to realize a fractional quantum Hall state \cite{leonard2023}.

In particular, the combination of periodic driving with a constant force that tilts the lattice potential has been instrumental for the implementation of artificial gauge fields. The constant force suppresses tunneling between lattice sites, while other mechanisms, such as laser-assisted tunneling \cite{jaksch2003, gerbier2010} or near resonant driving \cite{sias2008, haller2010}, reintroduce the coupling between lattice sites with the desired properties. These so-called Floquet-engineered lattice potentials have been successfully applied to investigate single-particle effects or weakly interacting quantum gases. Simulating many-body states of interacting particles, however, remains challenging as interactions create instabilities and heating on short timescales comparable to the modulation period and quickly destroy the system's coherence \cite{weitenberg2021c, bilitewski2015b}. Developing an understanding of these instabilities and finding an optimal window for the driving frequency \cite{sun2020} are crucial to include interactions in quantum-simulation experiments with periodic driving.

\begin{figure*}[t]
\includegraphics[width=0.95 \textwidth]{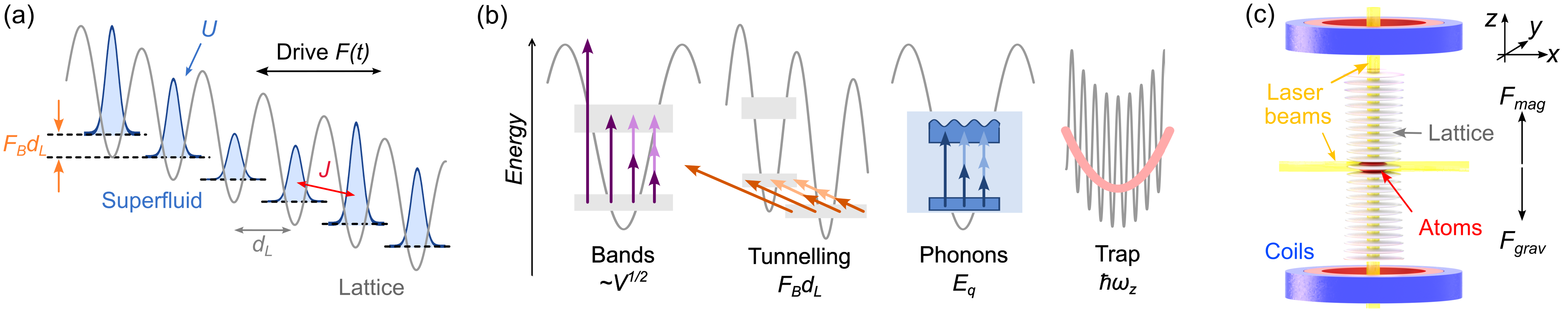}
\caption{Experimental setup and characteristic energies. (a) Sketch of a superfluid with density modulations (blue wave packets) in a tilted lattice potential (gray line) with tunneling energy $J$, interaction energy $U$, lattice spacing $d_L$, constant force $F_B$, and driving force $F(t)$. (b) Resonant excitations can occur when the driving energy $h \nu_D$ matches multiples or fractions of the energy of the band gap for lattice depth $V$, the tilt of lattice potential $F_B d_L$, phonon modes $E_q$, or trap frequencies $\hbar \omega_z$. Dark arrows show resonances of the driving frequencies and light arrows indicate their fractions for higher-order resonances. (c) Experimental setup with optical lattice, trapping laser beams, magnetic field coils, and levitating force.}\label{Fig:Setup}
\end{figure*}

In this study, we experimentally and theoretically investigate heating due to the spontaneous formation of phonon modes in a tilted lattice potential. We examine the role of interactions for the time evolution of a Bose-Einstein condensate (BEC) of cesium atoms confined in a 1D lattice potential that is periodically shaken and subject to a constant force, $F_B$ [Fig.\,\ref{Fig:Setup}(a)]. In such a system, phonon modes can grow exponentially in time and destroy the BEC \cite{sun2020}. Besides these phonon instabilities, the system exhibits a multitude of other driving resonances [Fig.\,\ref{Fig:Setup}(b)]. Resonant excitations to higher lattice bands occur when the driving frequency matches the energy gap between the bands or fractions thereof \cite{weinberg2015a, reitter2017}. In analogy to electrons in solids that are driven by electric fields, these higher-order excitations are often referred to as ``multiphoton-like'' \cite{arlinghaus2010a}. Furthermore, tunnelling resonances between neighboring lattice sites occur when the driving frequency matches the energy shift between sites. Multiples of this driving frequency couple sites at further distances while fractions of it allow for multiphoton-assisted tunneling \cite{eckardt2005a, sias2008, haller2010}. For tilted potentials, phonon instabilities coincide with these interband and tunnelling resonances and add another manifold of higher-order driving resonances to the system. Finding a frequency window free of resonances is experimentally challenging and requires a detailed understanding of the underlying mechanisms.

Within the lowest lattice band, periodically driven quantum gases show an exponential growth of phonon modes due to modulational and parametric instabilities \cite{dicarli2023}. Parametric instabilities are caused by oscillating system parameters  \cite{lellouch2017}, while modulational instabilities result from properties of the medium that exist even without periodic driving, such as attractive interactions or a negative effective mass for certain momenta of the superfluid in the lattice \cite{nguyen2017a,fallani2004}. They also occur in driven systems when the driving force accelerates the medium on a micromotion in momentum space. Parametric and modulational instabilities in superfluids were demonstrated independently in non-tilted lattice potentials for different driving regimes \cite{dicarli2023}. In the presence of a tilt of the potential, however, the superfluids' micromotion always covers the complete first Brillouin zone which makes it experimentally challenging to separate both mechanisms.

Floquet theories provide a very successful approach to analyze quantum gases in the fast driving limit when driving frequencies exceed all other frequencies of the system. In a high-frequency approximation, the system is well-described stroboscopically using a time-independent Floquet Hamiltonian \cite{eckardt2017}. Recently, this description was extended to lower frequencies to analyse parametric resonances \cite{bukov2015, lellouch2017, lellouch2018, boulier2019, wintersperger2020d}. Here, we take the opposite approach and extend the description of phonons and modulational instabilities in non-driven systems to the case of resonant periodic driving. Our approach offers an intuitive explanation of our measurement results, and the description remains valid in the limits of both slow and fast driving. For intermediate driving frequencies, resonant excitations are caused by the periodic growth of phonon modes whenever the micromotion passes through a modulationally unstable region of the Brillouin zone. Band excitations, which occur for very large driving frequencies \cite{reitter2017}, and trap excitations, which require long observation times \cite{mitchell2021}, have little impact on our results and we omit them in our discussion.

We experimentally study phonon modes in three settings: non-driven with a constant force $F_B$ (Sec.\,\ref{sec:nodriving}), resonantly driven with a frequency that matches the energy shift between adjacent lattice sites, $h\nu_D=F_B d_L$, at distance $d_L$ (Sec.\,\ref{sec:resdriving}), and off-resonantly driven with detuning $h\Delta \nu = F_B d_L - h \nu_D$ (Sec.\,\ref{sec:noneresdriving}). For all settings, the superfluid cycles through unstable regions of the first Brillouin zone which causes the growth of excitation modes. As a result, resonant growth of excitations does not depend directly on the driving frequency, $\nu_D$, but on the frequency $\nu_c$ of those crossings into critical regions. Time-averaging, e.g., to calculate the time-averaged Bogoliubov energy of a phonon mode within a Floquet description, results in the loss of information about the shape of the micromotion. Instead, we use the number of crossings and the time-intervals between them to determine $\nu_c$ and the resonance condition.

\section{Experimental setup}

Our starting point for the experiment is a BEC of approximately $50,000$ cesium (Cs) atoms in a vertical optical lattice potential [Fig.~\ref{Fig:Setup}(c)] that was created using two counter-propagating laser beams with wavelength $\lambda=1064$\,nm, lattice constant $d_L=\lambda/2$ and lattice momentum $k_L=\pi/d_L$ \cite{dicarli2019b}. Typical lattice depths, $V$, were between $2-14\,E_r$, where $E_r$ is the recoil energy of Cs atoms. The BEC was confined by a cross-beam optical dipole trap with trapping frequencies $\omega_{x,y,z}=2\pi\times(11,18,14)$\,Hz and levitated by a magnetic field gradient \cite{kraemer2004,dicarli2019a}. We controlled the atomic interaction strength by tuning the s-wave scattering length, $a_s$, with a magnetic Feshbach resonance, before loading the wave packet into the lattice potential in 150\,ms. The values of $a_s$ and $V$ were varied for different measurements to provide experimentally convenient time scales for the growth of excitation modes.

We created the constant force, $F_B$, and the driving force, $F(t)$, with different experimental techniques. The constant force was applied by reducing the magnetic levitation in $1.5$\,ms to a fraction of the gravitational force, which results in Bloch oscillations with period, $T_B = h/(F_B d_L)$, and frequency, $\nu_B = 1/T_B$, that are directly observable in the superfluid momentum distribution in the lab frame. The driving force, $F(t)=F_0 \cos(2\pi \nu_D t)$, was applied by periodically shifting the sites of the lattice using two acoustic-optical modulators that create a frequency difference between the laser beams \cite{niu1996, bendahan1996a}. This method provides precise control over driving frequency and strength. However, $F(t)$ is an inertial force in the lattice frame and the resulting micromotion is only indirectly detectable in the lab frame by measuring the superfluid's total momentum or the weight of the reciprocal lattice peaks in momentum profiles \cite{arimondo2012}. By combining the two techniques, we were able to directly measure the Bloch period in the lab frame while facilitating fast driving frequencies in the kHz regime.

After loading the superfluid into the lattice potential, we applied $F_B$ and $F(t)$ for a time $t$ that was adjusted to the closest multiple of the driving period to reduce effects of the micromotion on the final momentum distribution. To minimize the impact of the trapping potential, we kept $t$ smaller than the trap period \cite{mitchell2021}. The lattice potential was switched-off instantly or ramped-off in 1.2\,ms to determine the real momentum or the quasi-momentum distribution with absorption imaging after an expansion time of typically 75\,ms. For measurements with a clearly identifiable carrier wave packet, we applied Gaussian fits to determine the atom number in the carrier wave, $N_C$, and in excitation modes, $N_E=N_\text{tot}-N_C$, where $N_\text{tot}$ is the total atom number. This approach was challenging for measurements with a significant faction of atoms in excitation modes and we instead counted the number of atoms in fixed momentum intervals where we expected the phonon modes or the carrier wave (see \ref{ax:measurements}).

\section{Excitations without periodic driving \label{sec:nodriving}}

\begin{figure}[t] 
\includegraphics[width=\columnwidth]{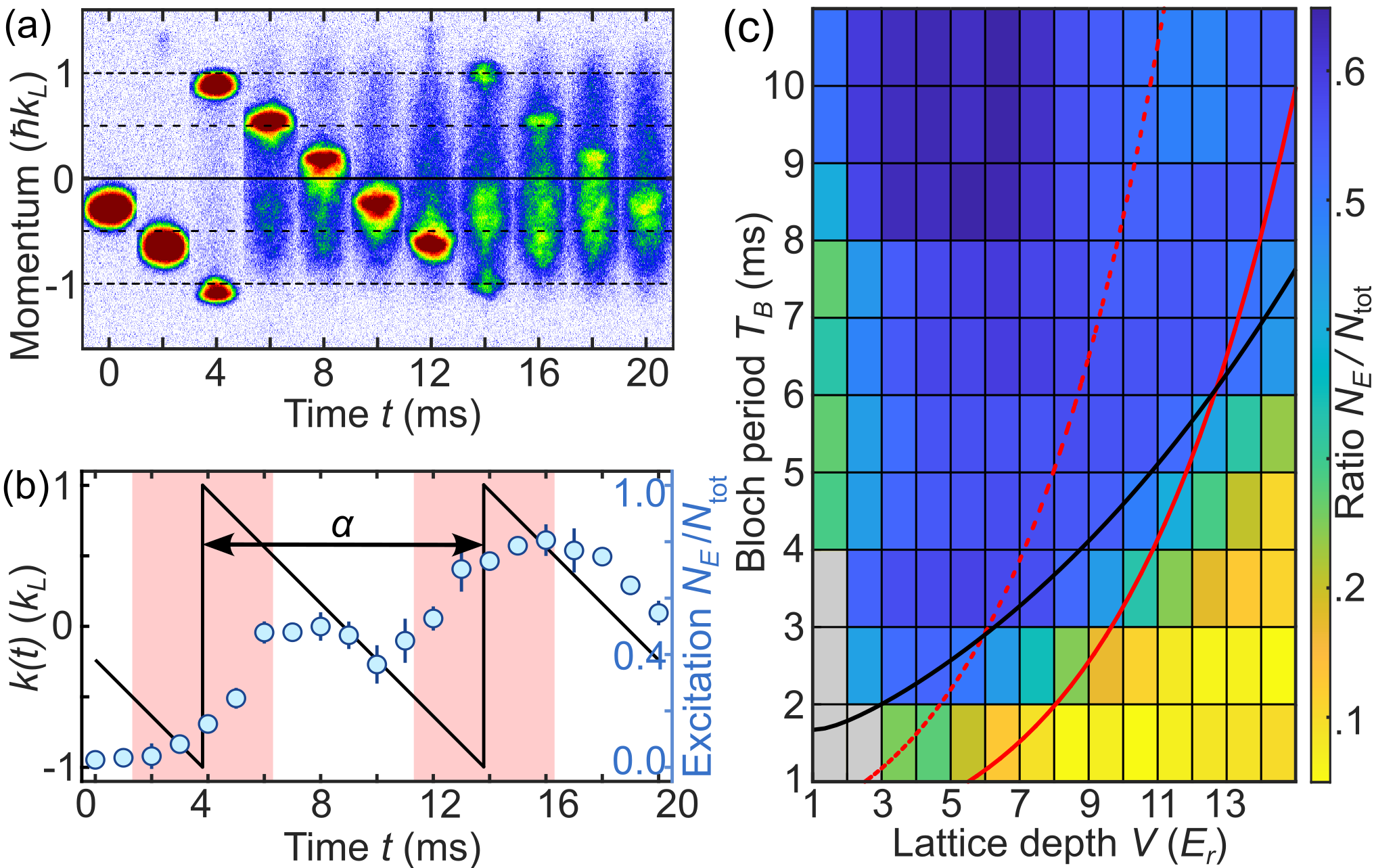}
\caption{Modulational instabilities without periodic driving. (a) Absorption images of Bloch-oscillating atoms and excitation modes in quasimomentum space for $T_B=10\,$ms, $V=6\, E_r$, $U/J=3$. (b) Atoms in excited modes $N_E/N_\text{tot}$ (blue circles) and calculated micromotion $k(t)$ (black line) for parameters in (a). Red patches indicate time intervals when micromotion crosses into unstable regions of the Brillouin zone. (c) Measured ratio $N_E/N_\text{tot}$ after a hold time of $t\approx20\,$ms. Lines indicate Bloch periods that match to the single particle bandwidth $4J/h$ (dashed red line), the approximated phonon energy $2E_{k_L}^s/h$ (solid red line), and the predicted transition line between unstable and stable parameters in \cite{zheng2004} (black line). Gray patches indicate strong atom loss for small $V$ and large values of $F_B$. }\label{Fig:ForceOnly} 
\end{figure}

To demonstrate the main concepts, we first studied excitation modes for only a constant force, $F_B$. Without driving, the superfluid Bloch oscillates in the first Brillouin zone with its micromotion, $k(t)$, following straight lines in quasimomentum space \cite{bendahan1996}. Absorption images, showing the momentum distribution in quasimomentum space [Fig.\,\ref{Fig:ForceOnly}(a)], were used to determine the fraction of atoms in excitation modes $N_E/N_\text{tot}$ [blue circles in Fig.\ref{Fig:ForceOnly}(b)]. Whenever the superfluid crosses into critical regions of the Brillouin zone with negative effective mass, $|k(t)|>k_L/2$, small perturbations of the stationary state grow exponentially in time \cite{zheng2004,fallani2004}, while excitations are steady or decay in stable regions \cite{dicarli2023} (see \ref{ax:theory}).

We quantified the system's stability for a varying constant force and lattice depth by measuring $N_E/N_\text{tot}$ after approximately $t=20$\,ms. Yellow and blue colors in the ($V,T_B$)-stability diagram [Fig.\,\ref{Fig:ForceOnly}(c)] indicate stable and unstable regions with weak and strong growth of phonon modes, respectively. We expect the superfluid to become stable in deep lattices and for large values of $F_B$ due to the suppression of tunneling when the energy gap between neighboring sites, $F_B d_L$, exceeds the width of the lattice band, $4J$, with tunneling matrix element $J$. However, we find that the decoupling of lattice sites is far less abrupt for a 1D-superfluid with large atom numbers per site than for single atoms \cite{ma2011}, and the superfluid remains phase coherent beyond $F_B d_L=4J$ [dashed red line in Fig.\,\ref{Fig:ForceOnly}(c)]. This allows us to observe the reduction of modulational instabilities when the system becomes stable in the fast cycling regime. This onset of stability shifts for increasing interaction strength towards smaller values of $T_B$.

For our analysis we assume that phonon modes with momentum $\hbar q$ are resonantly excited when the frequency $\nu_c$ of the carrier wave crossing into critical regions of the Brillouin zone matches twice the time-averaged Bogoliubov energy of the mode, $\brket{E_q}$. This resonance condition for modulational instabilities is similar to the condition for parametric instabilities  $h \nu_D = 2\brket{E_q}$ in references \cite{bukov2015c, lellouch2017, lellouch2018, wintersperger2020d}, but it relies on the frequency $\nu_c$. For strong driving forces, several crossings of critical regions can occur per Bloch period, and we introduce the parameter $\alpha$ to relate the resulting frequencies $\nu_c$ to $\nu_B$. The parameter $\alpha$ provides the fraction of time between two consecutive crossings per Bloch period with $\nu_c = \nu_B/\alpha$. As a result, our resonance condition predicts a series of values $T_B$ or $\nu_B$ with strong growth of excitation modes for
\begin{align} \label{eq:ResonanceCondition}
     h\nu_B = \frac{h}{T_B} \approx \frac{\alpha}{m_p}2\brket{E_q},
\end{align}
where $m_p$ is an integer that indicates higher-order phonon resonances (see Sec.\,\ref{sec:resdriving}).

The time-averaged energy of a phonon mode $\brket{E_q[k(t)]}$ with micromotion $k(t)$ can be approximated by two methods, both of which are based on the Bogoliubov-de-Gennes equations (BdG) \cite{hui2010, creffield2014, lellouch2017, lellouch2018}. Time-averaging the single-particle energy in the BdG equations provides an approximation, $E_q^f(K,k_0)$, that depends on the driving strength $K$ and on the initial momentum $k_0=k(0)$ \cite[\ref{ax:theory}]{zheng2004, creffield2009, hui2010, lellouch2018}. For fast driving frequencies, this analytic expression maps directly to the energy of a phonon $E_q(k)$ in a non-driven system when we identify $k$ with $k_0$ and the tunneling matrix element $J$ with $J_\text{eff}(K)$ (see \ref{ax:theory}). This mapping indicates that modulational instabilities also exist in periodically driven systems for $|k_0|>0.5 k_L$. However, the approximation assumes fast driving frequencies and neglects the short time intervals of phonon growth when the micromotion cycles through the Brillouin zone.

For a low-frequency approximation, we numerically time-average the energy $E_q[k(t)]$ of an existing phonon over $k(t)$ \cite{dicarli2023}
\begin{align} \label{eq:SlowDrivingLimit}
  E_q^s =  \frac{1}{T_B} \int_0^{T_B} E_{q}[k(t)] dt.
\end{align}
Due to modulational instabilities, the phonon energy $E_{q}(k)$ is a complex number with an imaginary component that indicates the mode's growth rate $\Gamma_q$ and a real component that provides the frequency of the mode's phase oscillations \cite{wu2003, modugno2004}. The same complex description applies to $E_q^s$, but not for the approximation $E_q^f$. Only the real component of $E_q^s$ is used in Eq.\,(\ref{eq:ResonanceCondition}) to predict resonant cycling frequencies.

Without driving, the micromotion crosses once per Bloch period into critical regions with $\alpha=1$ [Fig.\,\ref{Fig:ForceOnly}a] resulting in a periodic increase of the phonon mode occupation [Fig.\,\ref{Fig:ForceOnly}b]. Resonances occur when the frequency $\nu_c$ of these crossings matches to the natural oscillation frequency of the phonon density $2\brket{E_q}$ (with $\alpha=m_p=1$). For faster frequencies, the system becomes stable when $h \nu_c$ exceeds the energy of all phonon resonances, i.e., when the cycling due to Bloch oscillations becomes faster than possible response times of the phonons. We find that the transition to this fast cycling regime (or fast driving regime in later sections) is well approximated by Eq.\,(\ref {eq:ResonanceCondition}) [solid red line in Fig\,\ref{Fig:ForceOnly}(c)]. An even better prediction for the transition from an unstable to a stable system is provided by $T_B = h/(2.96\sqrt{2J U})$ [black line in Fig\,\ref{Fig:ForceOnly}(c)], which is an analytical approximation to a numerical solution of the Bogoliubov-de Gennes equations in \cite{zheng2004}.

We also observed a second stable region at shallow lattice depths $V<2\,E_r$ [Fig\,\ref{Fig:ForceOnly}(c)]. For repulsive interactions, modulational instabilities are induced by the lattice potential and we expect the superfluid to be become stable in the limit of vanishing lattice depths. We omit the discussion of instabilities in shallow lattice potentials as identifying unstable intervals of the micromotion is more challenging beyond the tight-binding regime \cite{wu2003}.

\section{Excitations with resonant driving \label{sec:resdriving}}

\subsection{Experimental measurements \label{sec:resdriveExp}}

\begin{figure}[t]
\includegraphics[width=\columnwidth]{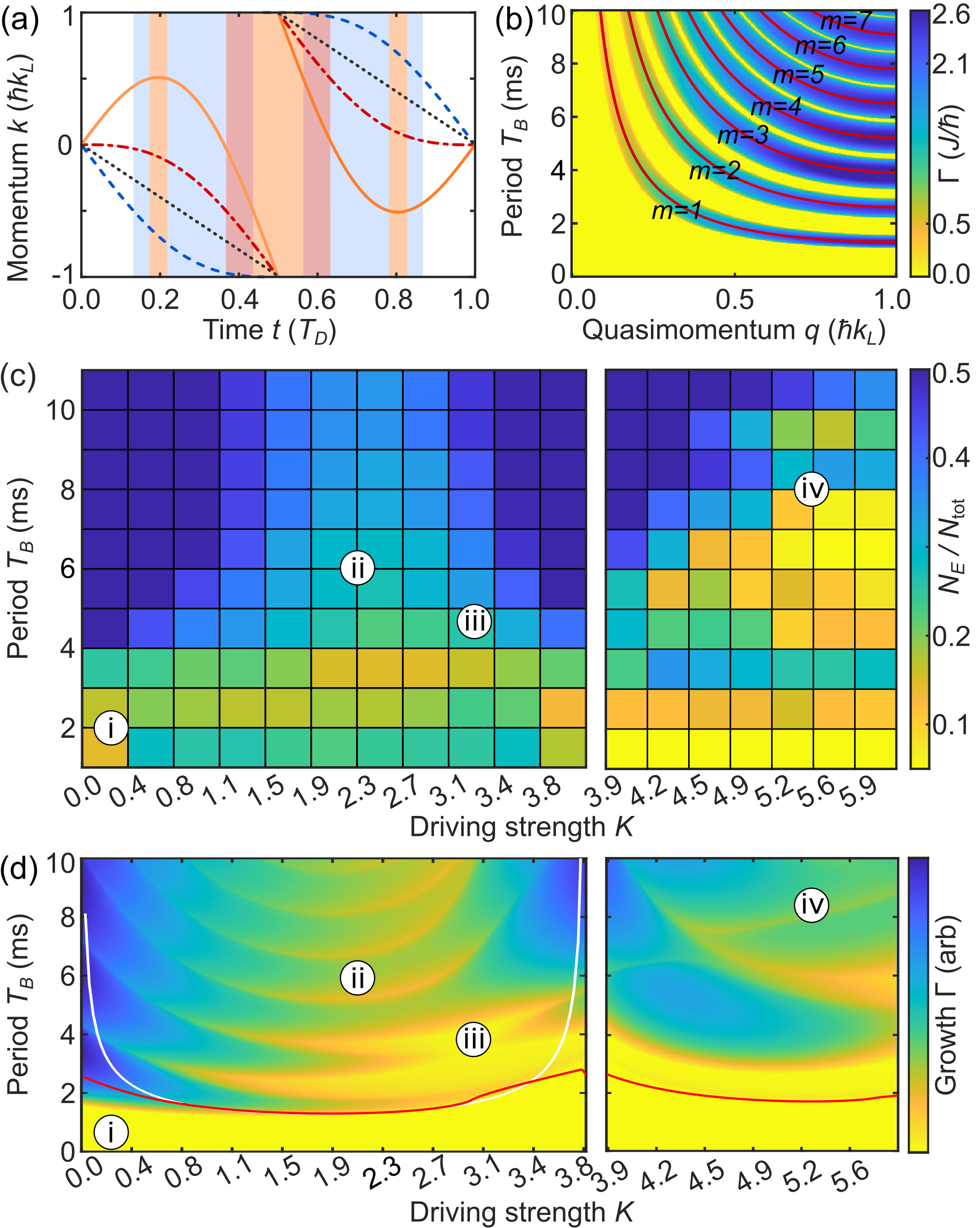}
\caption{Stability of a superfluid in a tilted potential with a resonant driving force. (a) Calculated micromotion for $K=0,1$ (black and blue lines) and for $K=1,3$ with opposite initial direction of $F(0)$ and $F_B$ (red and orange lines). Patches with corresponding colors indicate time intervals in critical regions of the Brillouin zone. (b) Numerical calculation of the growth rate $\Gamma$ for $K=2,V=10\,E_r,U/J=35$ together with predicted resonance (red line) with energy $2\Re[E_{k_L}^s]$ in Eq.\,(\ref{eq:ResonanceCondition}). (c) Experimental measurement of atoms in excitation modes after driving for approximately 30\,ms with strength $K$ and parameters $V=10\,E_r,N_\text{tot}\approx60,000,a_s=107\,a_0$. Left and right panels use initial momentum $k_0=0$ and $k_0=k_L$. Average over typically six repetitions. (d) Numerical calculation of $\Gamma$ with same parameters as in (c). Red and white lines indicate the predicted position of the fundamental resonance in Eq.\,(\ref{eq:ResonanceCondition}) when approximating the phonon energy by $E_q^s$ and $E_q^f$.}\label{Fig:MainStabilityDiagram} 
\end{figure}

In a second series of experiments, we added a resonant driving force $F(t)$ with a driving period that matches $T_B$. Despite a strongly tilted lattice potential, resonant driving re-introduces coupling between the lattice sites with an effective tunneling matrix element, $J_\text{eff}(K)=J J_1(K)$ \cite{eckardt2007}. Here, $J_1(K)$ is the first-order Bessel function and $K=F_B d_L/(h \nu_D)$ is the dimensionless driving strength. For increasing driving strength, the straight lines of the micromotion [$K=0$, dotted black line in Fig.\,\ref{Fig:MainStabilityDiagram}(a)] start bending upwards or downwards [red and blue lines in Fig.\,\ref{Fig:MainStabilityDiagram}(a)], depending on the initial directions of $F_B$ and $F(0)$. Colored patches in Fig.\,\ref{Fig:MainStabilityDiagram}(a) indicate time intervals during which the superfluid's micromotion crosses into critical regions of the Brillouin zone. We used opposite initial directions of the forces to reduce these time intervals [red and orange lines in Fig.\,\ref{Fig:MainStabilityDiagram}(a)], as identical initial directions extend them (blue dashed line).

To develop a better understanding of the system's stability, we calculated the growth rate, $\Gamma_q(T_B,K)$, of phonons with momentum $q$ by integrating and diagonalizing the Bogoliubov-de-Gennes equations \cite[\ref{ax:theory}]{creffield2009, hui2010, lellouch2018}. The resulting $(q,T_B)$-stability diagram for $K=2$ [Fig.\,\ref{Fig:MainStabilityDiagram}(b)] shows a sequence of unstable regions with strong growth (blue color) which belong to fundamental and higher-order phonon resonances $m_p=1-7$. The resonance condition Eq.\,(\ref{eq:ResonanceCondition}) provides a good approximation of these resonances (red lines with $\alpha=1$). Successive higher-order resonances are separated by a stable region (yellow color). Position, width, and growth rate for these stable and unstable regions can be related to the time that the micromotion spends in critical regions of the Brillouin zone [see \ref{ax:unstable}]. However, this $q$-dependence of $\Gamma$ is usually not directly visible in experiments due to the stochastic nature of modulational instabilities. Phonon modes require small initial modulations of the medium to start growing \cite{zheng2004}. Those seeds are of random nature, caused by thermal, quantum or technical fluctuations, which result in the occurrence of modes with varying values of $q$ \cite{tozzo2005a}.

We experimentally determined the growth of phonon modes by measuring $N_E/N_\text{tot}$ after a driving period of approximately 30\,ms. Two different initial quasimomenta, $\hbar k_0=0$ and $\hbar k_L$, were used to prepare the superfluid in the ground state of the time-averaged lattice band [panels in Fig.\,\ref{Fig:MainStabilityDiagram}(c)]. The ground state's micromotion starts with $k_0=0$ for $K<3.83$ when $J_\text{eff}(K)$ is positive and with $k_0=k_L$ for $3.83<K<7.02$ for negative values of $J_\text{eff}(K)$. To push the superfluid to the edge of the Brillouin zone, $k_0=k_L$, we applied a weak magnetic field gradient before starting the drive (see \ref{ax:measurements}).

We compare the measured ($T_B, K$)-stability diagram in Fig.\,\ref{Fig:MainStabilityDiagram}(c) to the calculated growth rate $\Gamma(T_B,K)$ averaged over 11 modes with equally spaced values of $q=0-k_L$ [Fig.\,\ref{Fig:MainStabilityDiagram}(d)]. Measurements and numerical calculations show good qualitative agreement for characteristic parameter regions that are indicated by Roman numerals in the diagrams. Particularly important for future experiments is the stable region at high driving frequencies [region i in Fig.\,\ref{Fig:MainStabilityDiagram}(d)] which, for weak driving strengths, starts approximately at the first resonance with $m_p=1$. Both $E_q^s$ and $E_{k_L}^f$ provide a good approximation for $\brket{E_q}$ in Eq.\,(\ref{eq:ResonanceCondition}), except for $K$ values close to the zeros of $J_1(K)$ [red and white lines in Fig.\,\ref{Fig:MainStabilityDiagram}(d)].

Stronger driving strengths, e.g.~$1.5<K<3.0$, shorten the time that the superfluid spends in critical regions and reduce growth rates and widths of the resonances. As a result, even large driving periods show an increased stability, both in our data and in calculations (region ii). Even stronger driving strengths, e.g.~with $K\approx 3$, cause three crossings of the micromotion into critical regions of the Brillouin zone. This reduces the $\alpha$ parameter and shifts the stable fast-driving regime towards larger values of $T_B$ (region iii). Even stronger driving strengths, $3.9<K<5.9$, for which the ground state has an initial momentum $k_0=k_L$, cause multiple crossings into critical regions with varying values of $\alpha$. Despite the resulting complex structure of stable and unstable regions in our calculation (region iv), we find good agreement between calculations and experiment within our measurement resolution.

\subsection{Interpretation of the stability diagram \label{sec:confirmAlpha}}

\begin{figure}[t]
\includegraphics[width=\columnwidth]{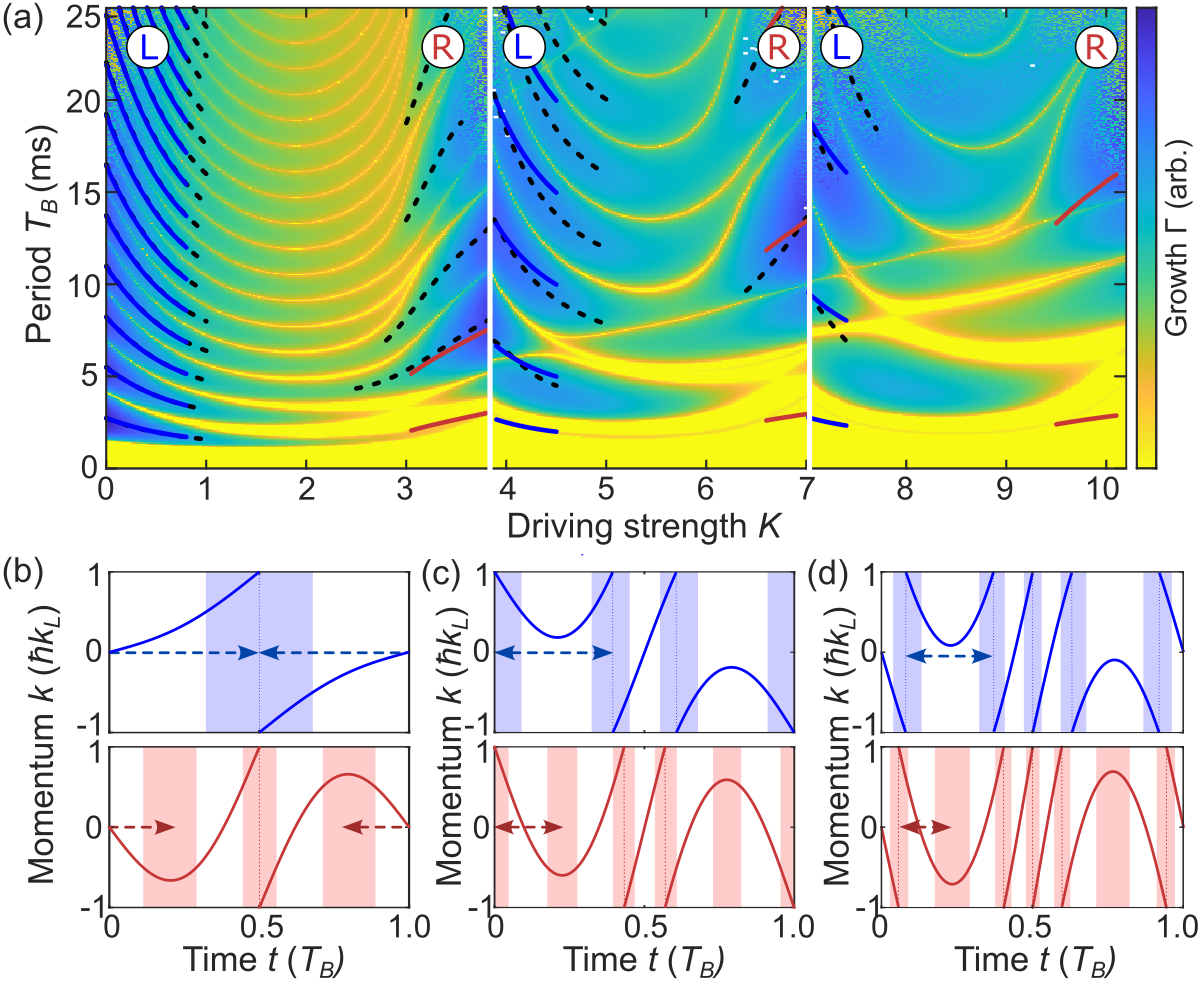}
\caption{Growth rate and driving resonances. (a) Calculated growth rate $\Gamma_{k_L}$ for $V=10\,E_r$ and $U=30\,J$. Dashed black lines indicate local maxima of $\Gamma_{k_L}$ that were determined with a peak-finding algorithm. Solid blue and red lines show the predicted resonance positions in Eq.\,(\ref{eq:ResonanceCondition}). Panels indicate different
signs of $J_\text{eff}(K)$. (b)-(d) Examples of the micromotion for the three panels in (a), top and bottom panels refer to regions (L) and (R) with blue and red line colors, respectively. (b) $K=0.5$ (top) $K=3.5$ (bottom), (c) $K=4.0$ (top) $6.5$ (bottom), (d) $K=7.5$ (top) $10.0$ (bottom). Colored patches indicate time intervals of the micromotion in regions of the Brillouin zone with modulational instabilities. Arrows show $\alpha$ for the longest time intervals between consecutive crossings.} \label{Fig:SimulationTilt}
\end{figure}

The structure of stable and unstable regions in our calculation of $\Gamma$ can be explained using Eq.\,(\ref{eq:ResonanceCondition}). For a direct comparison with Eq.\,(\ref{eq:ResonanceCondition}), we omitted the averaging and calculated $\Gamma_{q}(K,T_B)$ for a single mode, $q=k_L$, and increasing values of $K$ [Fig.\,\ref{Fig:SimulationTilt}(a)]. The local maxima of $\Gamma_{k_L}$ were determined using a peak finding algorithm (dashed black lines) and we compared them to the predicted resonances with parameters $(m_p,\alpha)$ (solid red and blue lines in Fig.\,\ref{Fig:SimulationTilt}).

We find that unstable regions on the left side of each panel [regions (L) in Fig.\,\ref{Fig:SimulationTilt}(a)] show a mostly regular pattern of higher-order resonances. The vertical distance between two resonances increases for each panel with $1/\alpha$ (blue lines) as predicted by Eq.\,(\ref{eq:ResonanceCondition}). $\alpha$ was determined using the micromotion for each value of $K$ [Figs.\,\ref{Fig:SimulationTilt}(b)-(d) for each panel in (a)]. The largest values of $\alpha$ [arrows in Figs.\,\ref{Fig:SimulationTilt}(b)-(d)] provide a good prediction for the resonance positions in regions (L). However, we find a more complex pattern of intersecting resonances on the right side of each panel in regions (R), and our prediction of the resonance position works less well [red lines in Fig.\,\ref{Fig:SimulationTilt}(a)]. The difference between regions (L) and (R) is caused by different number of crossings of the micromotion (colored patches), which have similar, but slightly different, time intervals between them in regions (R). We speculate that the structure of resonances is caused by this multitude of slightly different $\alpha$ values and their corresponding cycling frequencies $\nu_c$.

Equation (\ref{eq:ResonanceCondition}) provides a general condition for resonances in lattice systems with modulational instabilities and a cycling micromotion, irrespective of whether the cycling is caused by driving, a tilted lattice potential, or both. For systems without tilt, $\nu_B$ must be replaced by the driving frequency $\nu_D$, and $\alpha$ increases from 0 for weak driving strength, when the micromotion never crosses into critical regions of the Brillouin zone, to 1/2 for moderate driving strength. The resonance condition for this special case with $\alpha=1/2$ and $m_p=1$ is $h \nu_D = \brket{E_q}$ which matches the condition suggested in Ref.\,\cite{lellouch2017}.

\section{Excitations with non-resonant driving \label{sec:noneresdriving}}

\begin{figure}[t]
\includegraphics[width=\columnwidth]{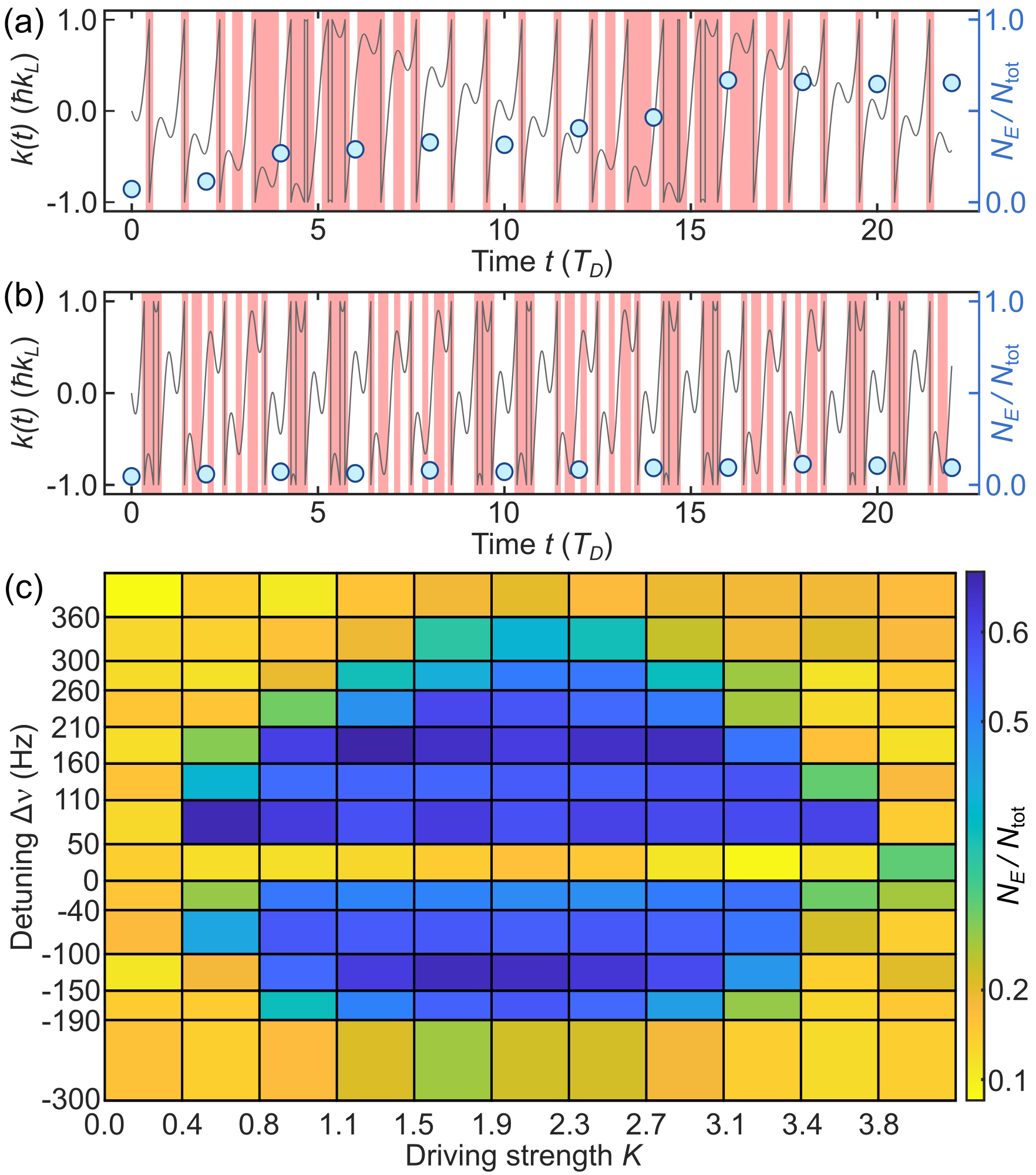}
\caption{Stability of the superfluid in a tilted potential with off-resonant driving. (a) Micromotion (gray line) and measured number of atoms in phonon modes $N_E$ (blue circles) for small detuning $\Delta \nu = 100\,$Hz with parameters $K=1.5,V=8\,E_r, k_0=0, T_B=0.998\,$ms, $U/J=15$, and (b) for large detuning $\Delta \nu = 800\,$Hz. Colored patches indicate time intervals in the critical region of the Brillouin zone. (c) Measured stability diagram showing $N_E/N_\text{tot}$ after approx.\,$30$\,ms of driving and parameters $V=8\,E_r,N_\text{tot}\approx 60,000, a_s = 107\,a_0$. }\label{Fig:Detuning} 
\end{figure}

Finally, we examine the stability for off-resonant driving frequencies with detuning $\Delta\nu = \nu_B - \nu_D$. Off-resonant driving in tilted lattice potentials can induce super-Bloch oscillations with large amplitudes in position space \cite{alberti2009,haller2010,kolovsky2010} that make the study of phonon modes challenging. To reduce the amplitude of those oscillations, we chose large detunings with $\Delta \nu>20$\,Hz and a strong force with $\nu_B=1$\,kHz. This Bloch frequency also ensures that the system is well in the fast-driving regime and any growth of phonon modes can be attributed to $\Delta \nu$. In momentum space, super-Bloch oscillations show the characteristic straight lines of Bloch oscillations in the Brillouin zone (as in Fig.\,\ref{Fig:ForceOnly}) when evaluating $k(t)$ stroboscopically at integer multiples of the driving frequency. The period of those oscillations is $\Delta T = 1/\Delta \nu$.

To understand the growth of phonon modes for non-resonant driving, it is helpful to study the complete micromotion $k(t)$. The shape of the micromotion gradually shifts, resulting in predominantly stable and predominantly unstable time intervals [red patches in Figs.\,\ref{Fig:Detuning}(a) and (b)], and it repeats itself after a period $T_\text{tot} = s/\nu_B=r/\nu_D$ for integer values $s,r$ and rational ratios $\nu_D/\nu_B$. The period $T_\text{tot}$ depends on the choice of $s,r$ and is less relevant for the phonon growth than the period between predominantly stable and unstable intervals, $\Delta T$. We found that the growth of phonon modes shows a similar time dependence for super-Bloch oscillations as for Bloch oscillations in Sec.\,\ref{sec:nodriving}. For instance, we observe for $\Delta T = 10$\,ms that excitation modes grow in the predominately unstable time intervals at 5\,ms and 15\,ms [blue circles in Fig.\,\ref{Fig:Detuning}(a)], and the system once again attains stability for larger detunings, e.g., for $\Delta \nu = 800$\,Hz [blue circles in Fig.\,\ref{Fig:Detuning}(b)].

As in Sec.\,\ref{sec:resdriving}, we quantified the growth of phonon modes by measuring $N_E/N_\text{tot}$ after approximately 30\,ms of driving. The resulting ($K,\Delta \nu$)-stability diagram shows a stable region for zero detuning [yellow color in Fig.\,\ref{Fig:Detuning}(c)], as expected in the fast-driving regime. Small values of $\Delta \nu$ cause instabilities as the superfluid spends longer time intervals in predominantly unstable regions (blue colors), while we find the system to be again stable for large detunings with $|\Delta \nu|> 360$\,Hz. The overall shape of the unstable region resembles an ellipse with a stable region in the form of a horizontal line at the center.

\begin{figure}[t]
\includegraphics[width=\columnwidth]{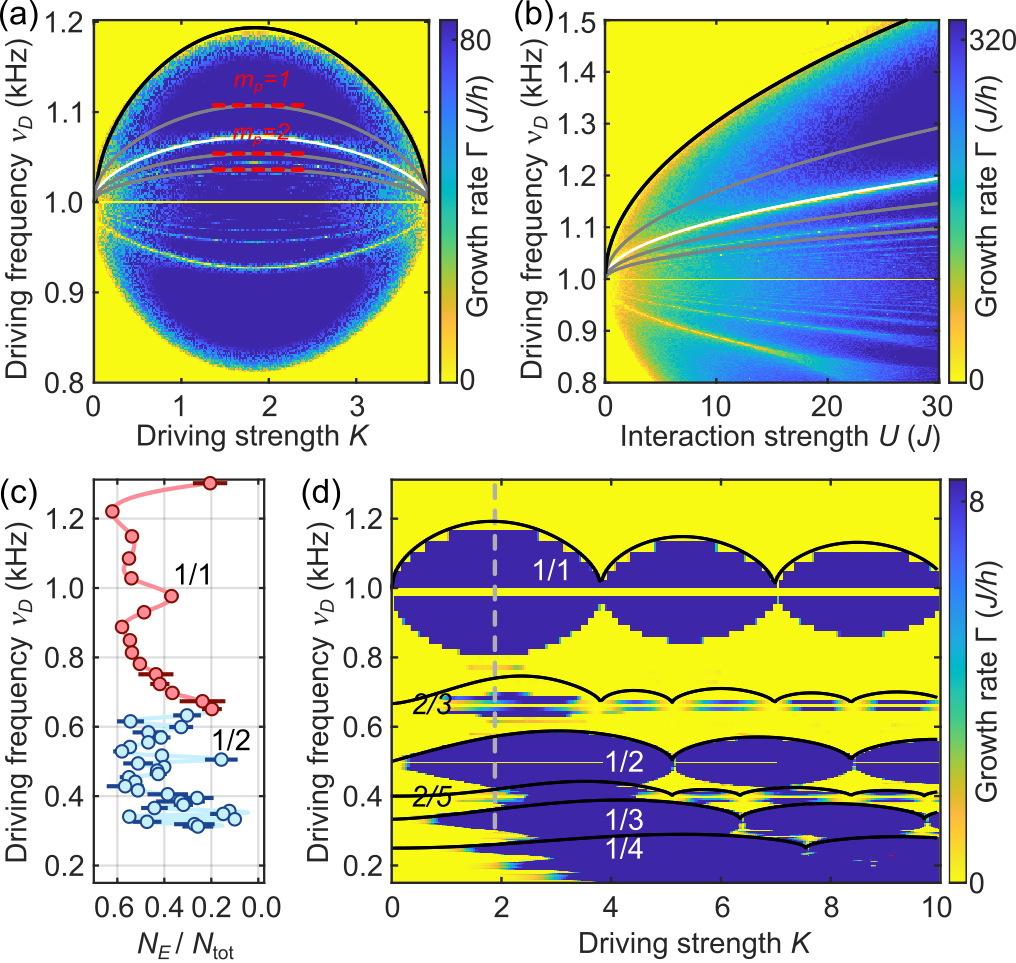}
\caption{Phonon growth rate for off-resonant driving. (a) Calculated growth rate $\Gamma_{k_L}$ for $V=10\,E_r$, $U=4J$, $k_0=0$, $s=500$, $\nu_B=1\,$kHz. Dashed red lines show the resonances $m_p=1,2,3$ for $K=1.85$ calculated using Eq.(\ref{eq:ResonanceConditionDetuned}). Other lines are based on the approximation $E_{k_L}^\text{tot}(K)$ in Eq.(\ref{eq:EnergyApproximiation}) predicting the resonances $m_p=1,2,3$ (gray lines), the stable region between $m_p=1$ and 2 (white line), and the transition between unstable and stable (black line). (b) Interaction dependence of $\Gamma_{k_L}$ for $K=1.85$. Other parameters and line colors match (a). (c) Measurement of  $N_E/N_\text{tot}$ for the parameters in Fig.\,\ref{Fig:Detuning}(c) and $K=1.90$ [dashed gray line in (d)]. Data colors indicate the tunneling resonances 1/1 (red color) and 1/2 (blue color). (d) Calculated growth rate for a large range of $K$ and $\nu_D$ values. Phonon resonances occur at tunneling resonances with frequency $\nu_D=(n_t/m_t)\nu_B$ and parameters $s=42$, $V=10\,E_r$, $U=4J$, $q=k_L$.}\label{Fig:SimulationDetuning}  
\end{figure}

We compared our measurement results to a numerical calculation of the phonon growth rate $\Gamma_q$ that was again based on the Bogoliubov-de Gennes equations. Instead of integrating the Bogoliubov-de Gennes equations over $T_B$, as in Sec.\,\ref{sec:resdriving}, we used the complete driving cycle $T_\text{tot}$. To match our experimental parameters and to provide rational values of $\nu_B/\nu_D$, we chose $\nu_B=1\,$kHz, $T_\text{tot} = sT_B$, and $T_D=(s/r)T_B$ with $s=500$ and integer values of $r$. We again used the phonon momentum $q=k_L$ that provides the strongest growth [Fig.\,\ref{Fig:SimulationDetuning}(a)]. The shape of the unstable region and the stable line at resonance match well to our experimental results. As for normal Bloch-oscillations in Fig.\,\ref{Fig:ForceOnly}(c), the transition lines between stable and unstable parameters regions are again well predicted by the approximation $h |\Delta \nu| = 2.96\sqrt{2J_\text{eff}(K) U}$ \cite{zheng2004}, where we replaced $J$ by the renormalized $J_\text{eff}(K)$ [black line in Fig.\,\ref{Fig:SimulationDetuning}(a)].

To explain the shape of the unstable regions in Fig.\,\ref{Fig:SimulationDetuning}(a), we use a resonance condition similar to the one in Eq.\,(\ref{eq:ResonanceCondition}). We replace the Bloch frequency with the super-Bloch frequency $\Delta \nu$ and use $\alpha=1$ for Bloch oscillations,
\begin{align} \label{eq:ResonanceConditionDetuned}
     h\Delta \nu \approx \frac{1}{m_p} 2\brket{E_q}.
\end{align}
Due to the detuning, the averaged phonon energy continues to oscillate when we time-average over the Bloch period, and we instead approximate $\brket{E_q}$ with a long time average $E_q^\text {tot}$ calculated over $T_\text{tot}$.
For a direct comparison, we use the momentum $q = k_L$ instead of averaging over different values of $q$. Equation\,(\ref{eq:ResonanceConditionDetuned}) provides a good approximation for the resonances [blue regions and dashed red lines in Fig.\,\ref{Fig:SimulationDetuning}(a)], however, $E_{q}^\text{tot}$ is independent of $K$ and our approach only predicts resonance positions close the maximum of $J_\text{eff}(K)$, i.e.~ at $K\approx 1.85$. There we recovered the full $K$-dependence of the resonance position by scaling the black transition line to $E_{k_L}^\text{tot}$, with
\begin{align} \label{eq:EnergyApproximiation}
     E_{k_L}^\text{tot}(K) = 1.64\sqrt{2J_\text{eff}(K) U}.
\end{align}
Gray lines in Fig.\,\ref{Fig:SimulationDetuning}(a) indicate the predicted positions of the phonon resonances with $m_p=1,2,3$. For completeness, we also compared the interaction dependence of the growth rate for $K=1.85$ to Eq.\,(\ref{eq:ResonanceConditionDetuned}) [Fig.\,\ref{Fig:SimulationDetuning}(b)] and found excellent agreement. The lines in Fig.\,\ref{Fig:SimulationDetuning}(b) refer to the same resonances as in Fig.\,\ref{Fig:SimulationDetuning}(a).

We extended our calculation of $\Gamma_{k_L}$ to larger driving strengths and detuning to demonstrate the connection between  phonon resonances and tunneling resonances [Fig.\,\ref{Fig:SimulationDetuning}(d)]. As in Figs.\,\ref{Fig:MainStabilityDiagram}(c) and (d), we used the momentum of the superfluid's ground state in the fast driving limit as the initial momentum $k_0$. The driving frequency $\nu_D$ is provided on the vertical axis instead of $\Delta \nu$ to allow for an easy recognition of tunneling resonances. For reference, the unstable region $m_p=1$ in Fig.\,\ref{Fig:SimulationDetuning}(a) matches to the first region at $\nu_D = 1\,$kHz in the top left corner of Fig.\,\ref{Fig:SimulationDetuning}(d). The unstable region is repeated for increasing $K$ values with a shape that is given by the Bessel function in $E_q^\text{tot}(K)$. Solid black lines indicate the transition lines between stable and unstable regions.

We found that phonon resonances also occur at other driving frequencies due to additional tunneling resonances [Fig.\,\ref{Fig:SimulationDetuning}(d)]. In strongly tilted lattice potentials, phonon resonances can only exist on top of a tunneling resonance, as phonons typically spread over several lattice sites which requires tunneling and coherence between the sites. Periodic driving restores the tunneling in tilted potentials when the driving frequency is resonant to the energy shift $h\nu_B$. For driving frequencies $\nu_D = (n_t/m_t)\nu_B$, the tunneling resonances are of order $m_t$ and couple lattice sites at a distance $n_t d_L$ \cite{sias2008, haller2010}. Combining all contributions, the complete resonance condition for phonon resonances of order $m_p$ and for tunneling resonance $(n_t/m_t)$ is
\begin{align} \label{eq:resonanceDetuned}
    h \nu_D =  \frac{n_t}{m_t} \left(h \nu_B + \frac{2\brket{E_q}}{m_p} \right).
\end{align}
Higher-order tunneling resonances require the regularization of $J$ with the $m_t$-order Bessel function $\JJ_{m_t}(K)$ \cite{kolovsky2010} which must be included when approximating the average phonon energy with $E_q^\text{tot}(K)$. We indicate the predicted driving frequencies for transition lines in Fig.\,\ref{Fig:SimulationDetuning}(d) with black lines and find good agreement with the unstable regions in our calculation. For the tunneling resonances with $n_t=2$, we used products of Bessel functions to provide a guide to the eye.

To experimentally demonstrate the existence of phonon resonances at the next tunneling resonance we measured $N_E/N_\text{tot}$ after 30\,ms of driving with a constant value $K=1.9$ [dashed gray line in Fig.\,\ref{Fig:SimulationDetuning}(d)]. We observe two regions with strong growth of phonon modes and stable center points at $\nu_D=1.0$\,kHz and $0.5$\,kHz [data points with red and blue colors Fig.\,\ref{Fig:SimulationDetuning}(d)]. The first region matches to our previous measurement in Fig.\,\ref{Fig:Detuning}(c) for the tunneling resonance $m_t=n_t=1$, while the second region occurs at tunneling resonance $m_t=2,n_t=1$. For our measurement parameters in Fig.\,\ref{Fig:Detuning}(c) and \ref{Fig:SimulationDetuning}(d), the unstable regions of phonon resonances start merging at higher-order tunneling resonances and we used a smaller interaction strength of $U=4J$ and a larger lattice depth $V=10\,E_r$ for the calculations to clearly show disjunct regions with phonon growth.


\section{Conclusion}

In conclusion, we have studied the growth of phonons in a superfluid with a tilted 1D lattice in three scenarios: non-driven, resonantly driven and non-resonantly driven. To determine the phonon growth, we measured the momentum distribution of the system after a fixed hold time and obtained the fraction of atoms that were not in the ground state. For all settings, we found stable and unstable parameter regimes which can be explained by analyzing the micromotion of the superfluid through the first Brillouin zone. Due to the tilted potential, the micromotion always crosses into critical regions of the Brillouin zone and modulational instabilities make the superfluid unstable for short time intervals. We found that the duration and multitude of those crossings per cycle determine the growth rate of phonon modes. Time-averaging over the micromotion, e.g.~within a Floquet-description, loses this information and makes it challenging to predict the stability of the system.

To determine the resonance condition, we matched the phonon energy with the frequency at which the micromotion crosses critical regions of the Brillouin zone. This cycling frequency allowed us to predict the fundamental and higher-order phonon modes of modulational instabilities in all three scenarios. For off-resonant driving, we replaced the driving frequency with the detuning $\Delta \nu$ between Bloch frequency and driving frequency. In all cases, a stable, fast-driving limit is reached when the cycling frequency exceeds the energy of the fundamental phonon mode. In addition to phonon resonances, band excitations and tunneling excitations, a complete stability analysis must include the role of the trapping potential \cite{mitchell2021}. Understanding the joint effects of these excitation mechanisms will be instrumental for quantum simulation experiments that study interacting many-body states with periodic driving over long time scales \cite{weitenberg2021c}.

\vspace{5ex}

We acknowledge support by the EPSRC through a New Investigator Grant (EP/T027789/1), the Programme Grant DesOEQ (EP/P009565/1), and the Quantum Technology Hub in Quantum Computing and Simulation (EP/T001062/1). CEC was supported by the Universidad Complutense de Madrid through Grant No. FEI-EU-19-12.

\renewcommand{\thefigure}{S\arabic{figure}}
\setcounter{figure}{0}

\renewcommand{\theequation}{S\arabic{equation}}
\setcounter{equation}{0}

\renewcommand\thesection{APPENDIX \Alph{section}}
\setcounter{section}{0}

\vspace{5ex}

\section*{Appendices}
\section{Description of excitation modes \label{ax:theory}}

This section summarizes the description of phonon modes in superfluids for lattice potentials with a constant force $F_B$ and a driving force, $F(t)=F_0 \cos(2\pi \nu_D t)$. These forces accelerate the superfluid on a periodic motion within the first Brillouin zone in momentum space. This micromotion $k(t)$ provides the basis for the subsequent analysis.

Without forces, $F_B=F(t)=0$, a superfluid with phonon excitations can be described by a carrier wave with quasimomentum $\hbar k$ and a weak perturbation $\delta \phi_k(z,t)$ \cite{modugno2004, trombettoni2006}
\begin{align}
\psi(z,t) &= e^{-i\mu_k t/\hbar} e^{ikz} [\phi_{k}(z) + \delta \phi_k(z,t)], \label{eq:perturbation}\\
\delta \phi_k(z,t) &= \sum_q u_{kq}(z) e^{i(qz - \omega_q(k)t)} + v^*_{kq}(z) e^{-i(qz-\omega_q(k)t)}, \label{eq:Excitation}
\end{align}
where $\mu_k$ is the chemical potential and $\phi_{k}$ is the solution of the stationary Gross-Pitaevskii equation. The perturbation is expressed as a superposition of Bogoliubov modes each with quasimomentum $\hbar q$, amplitudes $u_{kq},v_{kq}$, and energy $E_q=\hbar \omega_q(k)$. Due to momentum conservation those excitation modes occur in pairs of opposite quasimomentum $(+\hbar q,-\hbar q)$.

The energy of a single phonon mode with momentum $\hbar q$ and carrier momentum $\hbar k$ is given by \cite{wu2003,trombettoni2006}
\begin{multline} \label{eq:ExTrombettoni}
    E_{q}(k) = 2J \sin(k d_L)\sin(q d_L)  \pm 2 \\ \sqrt{ 4J^2 \cos^2(k d_L)\sin^4  \left( \frac{q d_L}{2}\right) + 2 J U  \cos(k d_L) \sin^2 \left(\frac{q d_L}{2}\right)},
\end{multline}
where $J$ and $U$ are the tunneling matrix element and the interaction energy. The real part of $E_{q}(k)$ describes the energy and the phase oscillations of the mode in Eq.\,(\ref{eq:Excitation}), while the imaginary part provides its growth rate $\Gamma = \Im[E_q]/h$. The largest value of $\Re[E_q(k)]$ for any combination of $q$ and $k$ in the Brillouin zone is $W = 2 \sqrt{ 4J^2 + 2 J U }$.

Adding a constant force $F_B$ suppresses tunneling when the energy shift, $F_B d_L$, between adjacent lattice sites is much larger than the single-particle band width, $4J$. However, resonant driving with frequency $h \nu_D = F_b d_L$, restores the coupling between lattice sites, and the wave packet shows a periodic micromotion through the first Brillouin zone
\begin{align}
    k(t) = k_0  +  \omega_D t - \frac{K}{d_L} \sin(\omega_D t + \varphi).
\end{align}
Here, $K=F_0 d_L/(\hbar\omega_D)$ is the dimensionless driving strength \cite{arlinghaus2011}. The phase $\varphi$ is set by the switch-on procedure of the forces and their initial directions. We used opposite directions for $F_B$ and $F(0)$ in calculations and experiments to minimize the growth of excitations for $k_0=0$.

An effective dispersion for the driven lattice system, $-2 J_\text{eff}(K) \cos(d_L k_0)$, can be derived by time-averaging the single-particle energy over one period of the micromotion $k(t)$ \cite{eckardt2007}. Here, $J_\text{eff}(K)=J J_1(K)$ is an effective tunneling element with the first-order Bessel function $\JJ_1(K)$. The sign of $J_\text{eff}(K)$  is negative in the interval $3.8<K<7.0$, resulting in an initial momentum $k_0=k_L$ for the ground state. We indicate this change of ground states with separate panels in Figs.\,\ref{Fig:MainStabilityDiagram}(c) and (d), and Fig.\,\ref{Fig:SimulationTilt}(a).

\begin{figure}[t]
\includegraphics[width=\columnwidth]{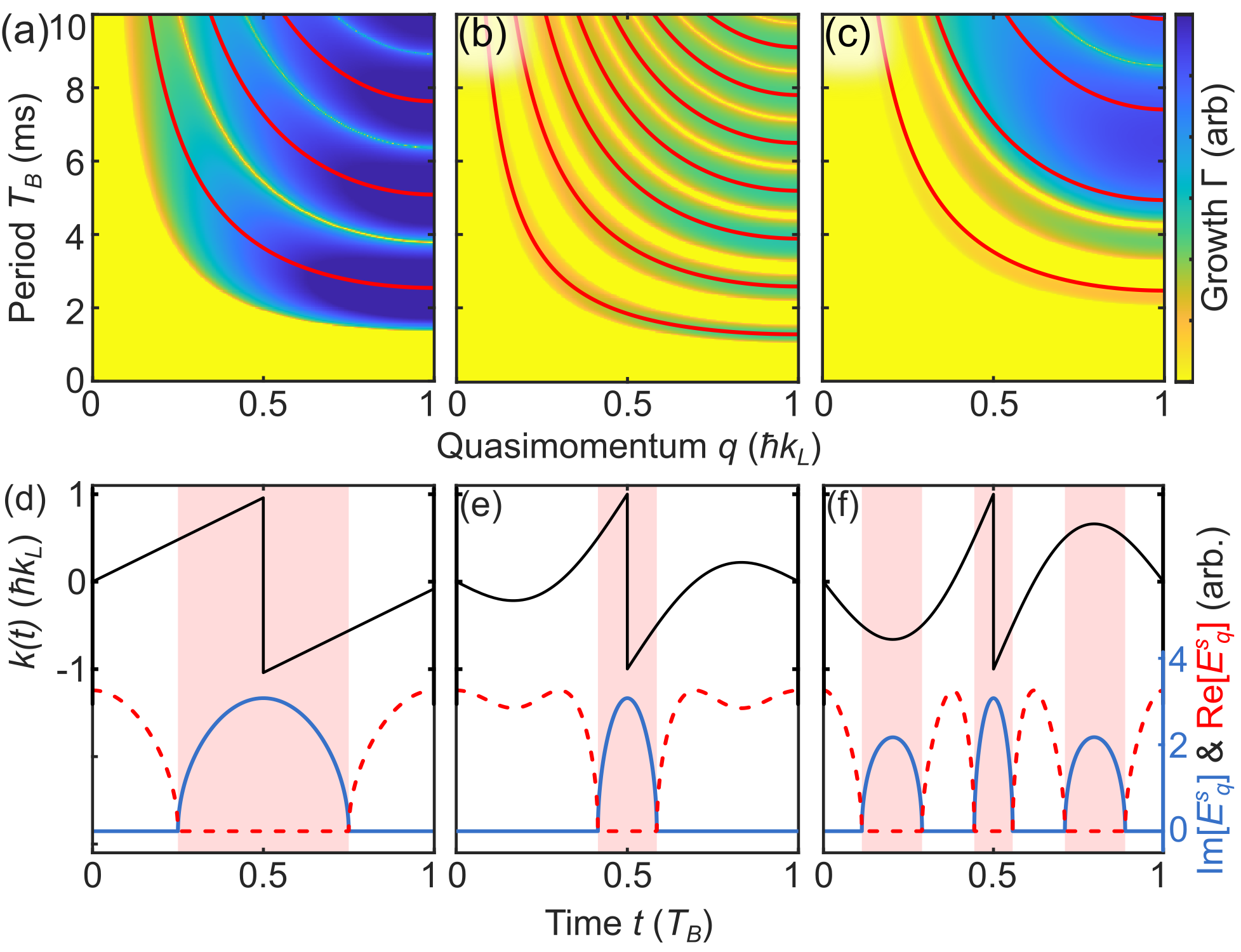}
\caption{Calculated growth rate for phonon modes with momentum $q$. (a)-(c) Growth $\Gamma_q$ for resonant driving $h \nu_D=F_B d_L=h\times1\,$kHz, $V=10\,E_r$, $U=35J$, and driving strength with (a) $K=0$, (b) $K=2$, (c) $K=3.5$. Red lines indicate Bloch periods for phonon resonances with energy $(2E^s_q)/m_p$ and integer values of $m_p$. (d)-(f) Micromotion $k(t)$ (black lines), phonon energy $\Re[E_{k_L}^s]$ (red lines), and growth rate $\Im[E^s_{k_L}]/h$ (blue lines) for (d) $K=0$, (e) $K=2$, (f) $K=3.5$.} \label{Fig:GrowthTheo}
\end{figure}

The Bogoliubov-de Gennes (BdG) equations provide the time evolution of a phonon mode \cite{zheng2004,creffield2009}. The components $u$ and $v$ of the mode in Eq.\,(\ref{eq:Excitation}) evolve according to
\begin{align} \label{eq:BogoliubovGennes}
 i\partial_t \begin{pmatrix} u_q \\ v_q \end{pmatrix} = \begin{pmatrix} \epsilon_+(q,t) + U & U \\ -U & - \epsilon_-(q,t)-U \end{pmatrix}  \begin{pmatrix} u_q \\ v_q \end{pmatrix},
\end{align}
where
\begin{multline}
     \epsilon_\pm(q,t) = 4 J \sin\left(\frac{q d_L}{2}\right)\\
      \times \sin\left(\frac{q d_L}{2}\pm k_0d_L \pm \omega_D t \mp K \sin(\omega_D t) \right). \nonumber
\end{multline}
In the limit of fast driving frequencies, $h\nu_D \gg W$, the BdG equations can be simplified by time-averaging $\epsilon_\pm(q,t)$ over one driving period and diagonalizing Eq.\,(\ref{eq:BogoliubovGennes}) \cite{creffield2009}. The resulting eigenvalues provide the energies of phonon and anti-phonon modes
\begin{multline}  \label{eq:FastDrivingLimit}
    E^f_q(k_0,K)= 2J_\text{eff} \sin(k_0 d_L)\sin(q d_L)  \pm 2 \\ {\small \sqrt{ 4J_\text{eff}^2 \cos^2(k_0 d_L)\sin^4  \left( \frac{q d_L}{2}\right) + 2 J_\text{eff} U  \cos(k_0 d_L) \sin^2 \left(\frac{q d_L}{2}\right)}.}
\end{multline}
We included $k_0$ in the description to extend the time-averaged Bogoliubov equation in \cite{lellouch2018}. As a result, Eq.\,(\ref{eq:FastDrivingLimit}) matches directly to the Bogoliubov dispersion relation for non-driven systems Eq.\,(\ref{eq:ExTrombettoni}), when replacing $J$ with $J_\text{eff}$, and $k$ with the initial momentum $k_0$. This condition is almost identical for a non-tilted system \cite{dicarli2023}, except that the first-order Bessel function in $J_\text{eff}$ is replaced by the zeroth-order.

For slow driving frequencies, $h\nu_D \lesssim W$, we time-average $E_{q}[k(t)]$ over one driving period [Eq.\,(\ref{eq:SlowDrivingLimit})]. $E^s_q$ provides the averaged energy of an existing phonon with momentum $q$ that follows a particular micromotion, while $E^f_q$ describes the energy of a phonon in a system with time-averaged parameters. Note that $E^f_q$ does not include the periodic growth of phonon modes during a driving cycle due to modulational instabilities which is the main subject of this article. The values of both energies, $\Re[E_q^s]$ and $E_q^f$, diverge close to the zeros of $J_1(K)$ but they agree well in between [red and white lines in Fig.\,\ref{Fig:MainStabilityDiagram}(d)].

\section{Calculation of the phonon's growth rate \label{ax:unstable}}

\begin{figure}[t]
\includegraphics[width=\columnwidth]{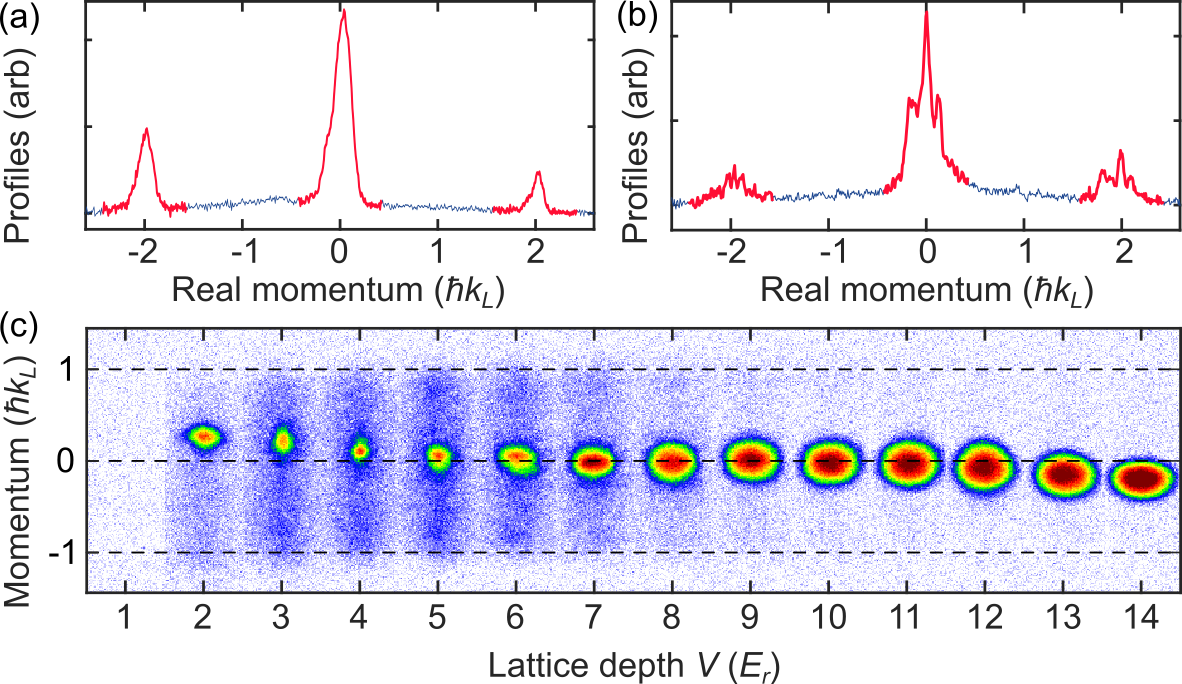}
\caption{Measurement of excitation modes. (a),(b) Integrated absorption profiles in real momentum space for (a) $K=0.8$, $T_B=4$\,ms and (b) $K=2.3$, $T_B=3$\,ms. Red lines indicate the regions used to determine the number of atoms in the carrier wave for Fig.\,\ref{Fig:MainStabilityDiagram}(c). (c) Absorption images for $T_B = 2$\,ms in Fig.\,\ref{Fig:ForceOnly}(c).} \label{Fig:MeasureFraction}
\end{figure}

We calculated the growth rate $\Gamma_q$ of phonon modes with momentum $q$ using Eq.\,(\ref{eq:BogoliubovGennes}) and the diagonalization procedure in Refs.~\cite{creffield2009, hui2010}. Figures \ref{Fig:GrowthTheo}(a)-(c) show $\Gamma_q$ for a driving strength $K=0,2,3.5$, with yellow and blue colors indicating stable and unstable parameter regions, respectively. Without driving ($K=0$) but with a constant force $F_B$, the micromotion passes only once per cycle through a critical region with non-zero growth [$\alpha=1$, Fig.\,\ref{Fig:GrowthTheo}(d)]. The corresponding time-averaged growth rate shows unstable regions that are boomerang-shaped and align with the predicted resonances in Eq.\,(\ref{eq:ResonanceCondition}) [Fig.\,\ref{Fig:GrowthTheo}(a)].

Adding a driving force, $K>0$, distorts the superfluid's micromotion [black line in Fig.\,\ref{Fig:GrowthTheo}(e)] and reduces the time it spends in critical regions of the Brillouin zone (red patch). This distortion of the micromotion changes the real and imaginary components of the time-averaged energy $E_q^s$. Spending less time in the critical region increases $\Re[E_q^s]$ as the real component of $E_q[k(t)]$ is zero in the critical regions, and it decreases $\Im[E_q^s]$ as the imaginary component is zero in the non-critical regions of the Brillouin zone. As a result, unstable regions shown in Fig.\,\ref{Fig:GrowthTheo}(b) move towards smaller values of $T_B$ and shrink in size compared to Fig.\,\ref{Fig:GrowthTheo}(a). This is the origin of the stable region (ii) in the $(K,T_B)$-stability diagram in Fig.\,\ref{Fig:MainStabilityDiagram}(d).

For larger driving strength, e.g.~$K=3.5$ in Figs.\,\ref{Fig:GrowthTheo}(c) and (f), the micromotion passes three times per cycle through a critical region. Due to these crossings, the relative time interval between two crossings, $\alpha$, decreases to $0.4$, and the cycling frequency $\nu_c$ increases. As a result, the fast driving limit shifts towards larger values of $T_B$ which is the cause for the stable region (iii) in Fig.\,\ref{Fig:MainStabilityDiagram}(d).

\section{Data acquisition and data analysis \label{ax:measurements} }

We determined the superfluid's distribution in real momentum and in quasimomentum space by taking absorption images either after a rapid switch-off or after a linear ramp of the lattice potential over $1.2$\,ms. We applied Gaussian fits to determine the atom number in the carrier wave, $N_C$, for data sets with a clearly identifiable carrier wave packet. This approach was challenging for measurements with a significant faction of atoms in excitation modes, and we instead counted the number of atoms in a momentum interval that enclosed the expected momentum of the carrier wave [red lines in Fig.\,\ref{Fig:MeasureFraction}(a)]. The fraction of atoms in excitation modes was calculated as $N_E/N_\text{tot} = 1 - N_C/N_\text{tot}$. Both methods do not account for phonon modes with small $q$ values [e.g.~in Fig.\,\ref{Fig:MeasureFraction}(b)], because the phonons and the carrier wave overlap in momentum space. However, excitation modes with large $q$ values dominate for strong interactions, $U>4J$, as used for the measurements in this article, and we expect that the omission of modes with small $q$ values does not change the overall shape of our stability diagrams.

\begin{figure}
\includegraphics[width=\columnwidth]{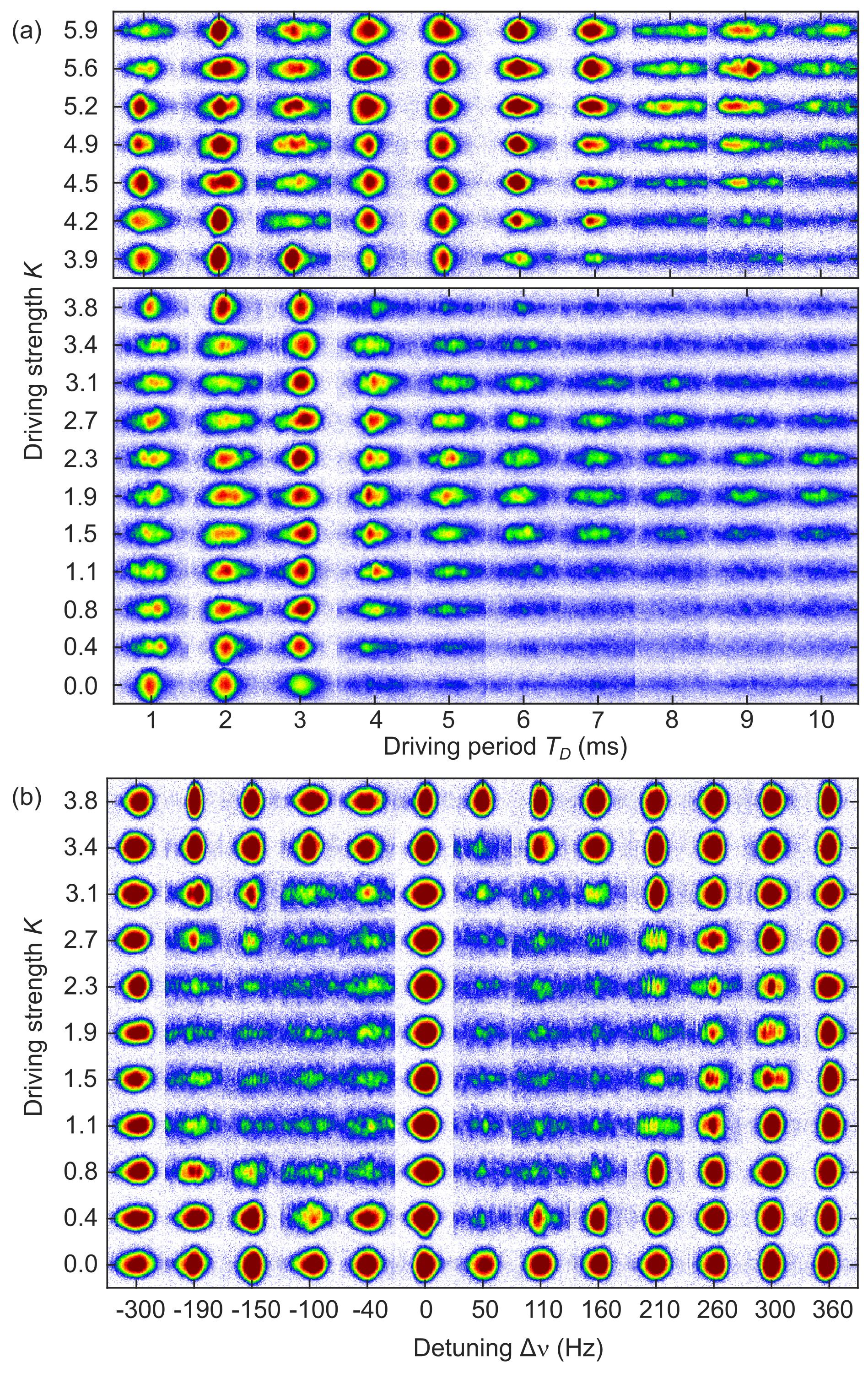}
\caption{(a) Absorption images used to generate the ($K,T_B$)-stability diagram with resonant driving in Fig.\,\ref{Fig:MainStabilityDiagram}. Only the area close the peak of the carrier wave is shown. Each image is the average of typically six absorption images with identical parameters. The top panel shows the sum of the two wave packets at $k_0=k_L$ for an easier comparison with the bottom panel with $k_0=0$. (b) Absorption images for ($K,\Delta \nu$)-stability diagram with off-resonant driving for Fig.\,\ref{Fig:Detuning}. The images show the area close to the central peak of the carrier wave. The panels are rotated by 90 deg.\,compared to Fig.\,\ref{Fig:MainStabilityDiagram} and Fig.\,\ref{Fig:Detuning}.} \label{Fig:AbsorptionImages}
\end{figure}

Changing the lattice depth $V$ during a measurement did sometimes introduce small additional forces that shifted the final momentum of the wave packet [e.g.~Fig.\,\ref{Fig:MeasureFraction}(c)]. Instead of balancing those small forces for every value of $V$, we included the final momentum shift of the wave packet in our data analysis when setting the momentum interval to determine $N_C$. For example, the scan of $V$ in Fig.\,\ref{Fig:MeasureFraction}(c) created a total variation of approximately $1\%$ of $F_B$, which we expect to have little influence on the stability measurement.

Figure \ref{Fig:AbsorptionImages} shows the averaged absorption images for the measurement of the ($K,T_B$)-stability diagram [Fig.\,\ref{Fig:MainStabilityDiagram}(c) in main text]. The system was always prepared in its ground state for the fast driving limit, i.e., with $k_0=0$ for $K=0-3.8$ [bottom panel in Fig.\,\ref{Fig:AbsorptionImages}(a)] and with $k_0=k_L$ for $K=3.9-5.9$ [top panel in Fig.\,\ref{Fig:AbsorptionImages}(a)]. For an easy comparison of both parameter regimes, we show only the peak of the carrier wave in the bottom panel and the sum of the two wave packets at the edge of the Brillouin zone in the top panel [Fig.\,\ref{Fig:AbsorptionImages}(a)]. Figure \ref{Fig:AbsorptionImages}(b) provides averaged absorption images for non-resonant driving, [Fig.\,\ref{Fig:Detuning}(c) in main text]. Again, we show only the momentum interval for the main peak of the carrier wave for reference.

\vfill

\bibliography{./Modulation_Gradient, ./OurPapers }

\end{document}